\documentstyle[aps,prl,twocolumn,psfig,epsf]{revtex}
\begin{document}
\flushbottom
\draft
\twocolumn[\hsize\textwidth\columnwidth\hsize\csname
@twocolumnfalse\endcsname
\title{Electron-hole liquid in the hexaborides}
\author{M. E. Zhitomirsky and T. M. Rice}
\address{
Institut f\"ur Theoretische Physik,
ETH-H\"onggerberg, CH-8093 Zurich, 
Switzerland 
}
\date{October 18, 1999}
\maketitle
\begin{abstract}
\hspace*{2mm}
We investigate the energetics of the electron-hole liquid in 
stoichiometric divalent metal hexaborides. The ground state energy
of an electron-hole plasma is calculated using RPA and Hubbard
schemes and compared to the binding energy of a single exciton.
Intervalley scattering processes play an important role in
increasing this binding energy and stabilizing a dilute Bose gas
of excitons. 
\end{abstract}
\vspace{5mm}
]
\narrowtext

The remarkable discovery of the high-temperature weak 
ferromagnetism in La-doped CaB$_6$, SrB$_6$, and BaB$_6$ 
has opened a new page in the physics of magnetism \cite{Young}. 
In our previous work \cite{cond} we attributed this effect 
to an unusual ground state of undoped divalent hexaborides. 
This so-called excitonic insulator is characterized by a 
condensation of bound electron-hole pairs (excitons).
An excitonic instability in narrow gap semiconductors or semimetals
was predicted \cite{Keldysh65} and studied theoretically
in the mid-sixties \cite{Halperin}.
However, its occurrence in any real compound is still controversial.
Band structure calculations \cite{Hasegawa,Massida} predict 
a small direct overlap in divalent-metal hexaborides between 
a boron-derived valence band and a cation-derived conduction band 
at three equivalent $X$-points in the cubic Brillouin zone. This 
feature, together with absence of direct electric-dipole transitions
between the two bands, is extremely favorable for electron-hole
pairing and leads to the excitonic instability. Weak ferromagnetism 
develops, then, in a triplet excitonic insulator due to spontaneous 
time-reversal symmetry breaking under doping \cite{Volkov,cond}.

There are many open questions in the physics of the excitonic 
ferromagnetism. Some of them were raised recently in 
Refs.~\cite{Balents,Gorkov}. In the present work we, however, want 
to shift emphasis from the (un)doped excitonic state to general
properties of electron-hole ($e$-$h$) liquids in the hexaborides. 
Charge conservation does not fix the number of $e$-$h$ pairs, which in  
thermal equilibrium depends on the $e$-$h$ chemical potential (band 
overlap). In the original theory of the excitonic insulator the band 
gap (overlap) $E_G>0$ ($E_G<0$) was considered to be a free parameter 
that changes continuously through the value $E_G=0$. Subsequently, 
investigation of optically pumped electrons and holes in semiconductors 
showed \cite{Keldysh68,Brinkman,Rice} that a first order transition 
between two states, one with a substantial band gap and the other with 
a substantial band overlap, will occur and that smaller values of 
$|E_G|$ lie in the unphysical intermediate region. The details of such 
transition are very sensitive to the actual band degeneracies and 
anisotropies. Here, we study possible scenarios for the transition from 
a semiconducting state into a metallic $e$-$h$ liquid in CaB$_6$, 
including the appearance of a free exciton gas. We find that a novel 
mechanism of intervalley scattering of excitons is important in 
stabilizing an intermediate gas phase.

\begin{figure}[hp]
\centerline{\psfig{figure=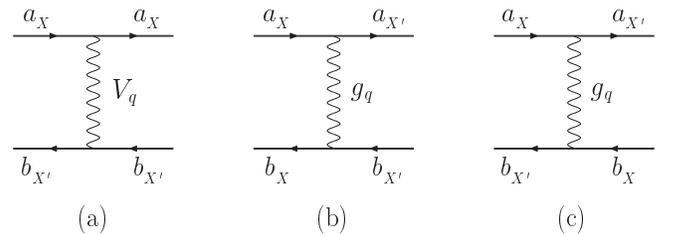,width=12.7cm,angle=0}}
\vspace*{-12.1cm}
\caption{Electron-hole scattering processes in the hexaborides. 
Operators $a$ and $b$ correspond to conduction and valence bands, 
respectively. Figure (a) shows a dominant intravalley vertex. 
Figures (b) and (c) correspond to intervalley scattering.}
\label{vertex}
\end{figure}

Electrons and holes in CaB$_6$ have the following values of the 
effective masses measured in units of the bare electron mass \cite{cond}:  
$m_e^\parallel=0.504$, $m_e^\perp =0.212$ (conduction band) and 
$m_h^\parallel = 2.17$,  $m_h^\perp = 0.206$ (valence band). These values 
agree with the early results using the muffin-tin approximation 
\cite{Hasegawa}. First, we retain only the dominant intravalley scattering 
processes shown in Fig.~\ref{vertex}a, when an electron (hole) scatters 
between states near a single $X$-point. In the small-$q$ limit the 
scattering matrix element is given by a screened Coulomb potential 
$V_q=4\pi e^2/\kappa q^2$ ($\kappa$ is a static dielectric constant). We 
define an $e$-$h$ chemical potential as a sum of two individual potentials 
$\mu = \mu_e + \mu_h$. The relation to the band model is made by 
$\mu = -E_G$, i.e.\ the band gap is equivalent to the chemical potential. 
Natural units for energies and lengths are effective Rydberg $E_x$ and 
Bohr radius $a_x$: $E_x = m^*e^4/2\kappa^2 = e^2/2\kappa a_x$, where 
reduced mass $m^*=m_{oe}m_{oh}/(m_{oe}+m_{oh})$ is determined by optical 
masses: $3/m_o = 2/m^\perp + 1/m^\parallel$. The $e$-$h$ pair density $n$ 
is characterized by a dimensionless parameter, 
$r_s = (3/4\pi na_x^3)^{1/3}$.

\vspace*{-2.5cm}
\vspace*{-3mm}

\begin{figure}
\unitlength1cm
\epsfxsize=8cm
\begin{picture}(2,3)
\put(0.1,-3){\epsffile{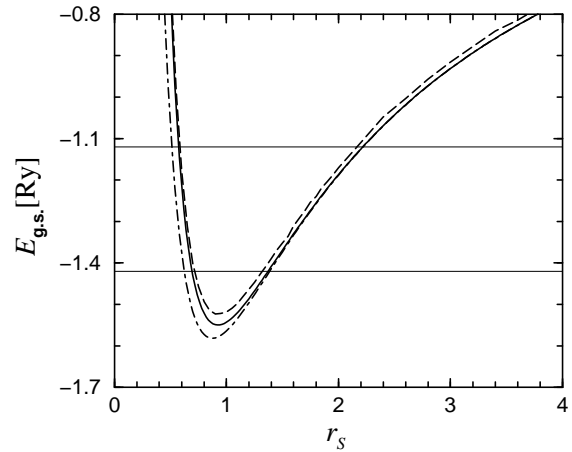}}
\end{picture}
\vspace{3cm}
\caption{The ground state energy per pair in units of $E_x$ for $e$-$h$ 
liquid in CaB$_6$. Solid line is the RPA and dashed line is the Hubbard 
result. Dot-dashed line is the RPA including correction from the intervalley 
scattering. Thin horizontal lines denote position of 
the $A_1$ exciton level in 
zeroth (upper) and first (lower) order approximations in the intervalley 
scattering. Self-consistent mechanism proposed in the text pushes the 
exciton energy further down below the local minimum.}
\label{egs}
\end{figure}

At high densities, small $r_s$, we use the random phase 
approximation (RPA) \cite{Brinkman}. Strictly speaking, a dense metallic 
$e$-$h$ liquid can transform into an excitonic insulator 
\cite{Keldysh65,Halperin}. This assumption was a key starting point in our 
explanation of the high-temperature weak ferromagnetism in the hexaborides 
\cite{cond}. However, the ground state energy correction from the 
excitonic instability in a semimetal is of order $\Delta^2/\varepsilon_F$ 
and is small ($\Delta\ll\varepsilon_F$), so we neglect it. The kinetic 
energy per $e$-$h$ pair is given by
\begin{equation}
E_K = \frac{3m^*}{5\alpha r_s^2\nu^{2/3}}
\left[\frac{1}{(m_e^\parallel m_e^{\perp2})^{1/3}}
+ \frac{1}{(m_h^\parallel m_h^{\perp2})^{1/3}} \right],
\end{equation}
where $\alpha = (4/9\pi)^{1/3}$ and $\nu$ ($=3$) is the number of valleys.
The exchange energy for anisotropic bands is
\begin{eqnarray}
E_{\rm exch} & = & - \frac{3}{2\pi\alpha r_s\nu^{1/3}}
\left[\phi(m_e^\perp/m_e^\parallel)+\phi(m_h^\perp/m_h^\parallel)\right],
\nonumber \\
 & & \ \phi(x) = x^{1/6}\,\frac{\arcsin \sqrt{1-x}}{\sqrt{1-x}}.
\end{eqnarray}
The correlation energy describes the remaining contribution to 
$E_{\rm g.s.}$
\begin{equation}
E_c \! = \! \frac{i}{2}\!  \int\!  \frac{dqd\omega}{(2\pi)^4} \! 
\int_0^1\!\! d\lambda \! 
\left[ \frac{V_q\Pi^*(q,\omega)}
{1\! -\!  \lambda V_q \Pi^*(q,\omega)}
\!-\! V_q \Pi^0(q,\omega) 
\right]\!\!,
\label{Ecorr}
\end{equation}
where $\Pi^*(q,\omega)$ is the irreducible polarization operator and
$\Pi^0(q,\omega) =\sum_i^{2\nu} \Pi^0_i(q,\omega)$ is a sum of 
(anisotropic) RPA-polarizabilities for each species of electrons or holes. 
Substitution $\Pi^*(q,\omega)=\Pi^0(q,\omega)$ in Eq.~(\ref{Ecorr}) gives 
the RPA-expression for $E_c$. An approximate way of treating the higher 
order exchange corrections was considered by Hubbard \cite{Hubbard}. His 
expression generalized to the multicomponent plasma is
\begin{equation}
\Pi^*(q,\omega) = \sum_i^{2\nu} \frac{\Pi^0_i(q,\omega)}
{1+f(q)V_q\Pi^0_i(q,\omega)} \ , 
\end{equation}
with $f(q) =0.5q^2/(q^2+k_F^2)$. We also change an $\omega$-integration 
from real to imaginary axis, which avoids a difficulty related to a plasmon 
pole in $\Pi^0(q,\omega)$. Numerical results for the ground state energy 
are shown in Fig.~\ref{egs}. The band degeneracy improves substantially 
the accuracy of the RPA, because corrections to the RPA diagrams have 
extra smallness in $1/2\nu$. The minimum of the ground state energy is 
reached at $r_s=0.92$ with a minimum value 
$E^{\rm min}_{\rm g.s.}=-1.55E_x$($-1.51E_x$) in the RPA (Hubbard) scheme. 
The use of RPA is justified by a small value of $r_s$ at the minimum, 
which corresponds to a dense plasma with strong screening and small 
corrections from multiple $e$-$h$ scattering. Such corrections become 
significant at $r_s\gtrsim 3$. The ground state energy is also known in 
the limit $r_s\rightarrow\infty$, where it approaches the binding energy 
of a single exciton $E_e$. (Here, we disregard possible formation of 
exciton molecules.)

\begin{figure}
\unitlength1cm
\epsfxsize=7.5cm
\begin{picture}(2,3)
\put(0.2,-3){\epsffile{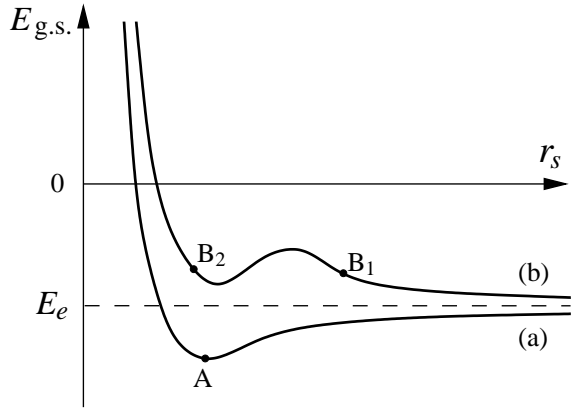}}
\end{picture}
\vspace{3.7cm}
\caption{Schematic form of the density dependence of the ground-state 
energy per pair in the two scenarios for metal-insulator transition in 
a semiconductor.}
\label{egs(rs)}
\end{figure}

The presence of a local minimum at metallic densities for
$E_{\rm g.s.}(r_s)$ has an important effect on transformation from 
a semiconductor to a semimetal \cite{Keldysh68,Brinkman,Rice}. 
The two possible scenarios are shown schematically in Fig.~\ref{egs(rs)}.
In the first case, curve (a), the local minimum is also the absolute one: $E_{\rm g.s.}(r_{sA})<E_e$. The pair chemical 
potential is related to the ground state energy by:
$\mu = E_{\rm g.s.} + n\frac{\partial E_{\rm g.s.}}{\partial n}$. At the 
extremal point the second term is zero and $\mu = E_{\rm g.s.}^{\rm min}$. 
Therefore, when the band gap decreases to $E_G=|E_{\rm g.s.}^{\rm min}|$,
a first-order metal-insulator transition takes place. The number of 
$e$-$h$ pairs jumps from zero to $n(r_{sA})$. Smaller densities with 
$r_s>r_{sA}$ correspond to unstable states. In optically pumped 
semiconductors, where a number of carriers is fixed instead of a chemical 
potential, this effect leads to $e$-$h$ droplet condensation \cite{Rice}. 
In the second case, curve (b), the exciton energy lies below the metallic 
minimum. Semiconducting state becomes unstable at $E_G=|E_e|$ and 
transforms into a low-density exciton gas. If $E_G$ is further reduced, a 
first order transition of gas-liquid type takes place between two states 
$B_1$ and $B_2$, which have same pressure 
$P=n^2\frac{\partial E_{\rm g.s.}}{\partial n}$. The case (b) is believed 
to be realized in isotropic one-component $e$-$h$ plasma, whereas in many 
semiconductors band degeneracies and anisotropies favor the case (a) 
bypassing a free exciton gas state \cite{Brinkman,Rice}.

To check which of the two scenarios occurs in the hexaborides we now compare 
the above value of $E^{\rm min}_{\rm g.s.}$ to the binding energy of a single 
exciton. The simplest estimate for the exciton energy is $-E_x$. It is obtained 
by substituting an isotropic $1s$-type wave function into the Schr\"odinger 
equation. This estimate predicts the same binding energy for excitons formed by 
an electron and a hole from same or from different valleys. We improve 
this result by using a simple variational ansatz appropriate to the 
cylindrical symmetry of the bands near each of the $X$-points similar to 
the treatment of shallow impurity states \cite{Kohn1}. The binding energy 
of an electron and a hole from the same valley is 
$E_e = -1.12E_x$ ($\langle a^\dagger_{X}b^{_{}}_{X}\rangle\neq 0$),
whereas an electron and a hole from different valleys have $E'_e = -1.02E_x$ 
($\langle a^\dagger_{X}b^{_{}}_{X'}\rangle\neq 0$). The former case with a more 
anisotropic reduced mass tensor is favored since lower dimensionality gives, 
as usual, a stronger binding. Thus, band anisotropy favors a particular 
type of excitons diagonal in the valley index. The chemical potential of the 
metallic $e$-$h$ liquid is below the exciton binding energy in this 
approximation. Hence, a dilute exciton gas is unstable and scenario 
(a) is predicted instead. 

The dominant term approximation (Fig.~\ref{vertex}a) used above 
leaves threefold degeneracy of the excitonic level corresponding to 
$e$-$h$ pairing at three $X$-points. Coherence between different excitons 
is established by intervalley scattering processes shown in 
Fig.~\ref{vertex}b. Instead of a single hydrogenic 
equation, one must solve now a system of three coupled integral equations 
for a three-component excitonic wave function $\psi_\nu(r)$. The scattering 
matrix element $g_q$ includes a large momentum transfer on $Q=b/\sqrt{2}$, 
where $b$ is an elementary reciprocal lattice vector. Consequently, we can 
neglect its momentum dependence and estimate 
\begin{equation}
g \approx \sum_G \frac{4\pi e^2}{|Q+G|^2}\:F^{aa}_G(X,X') F^{bb}_{-G}(X',X)\ .
\label{gq}
\end{equation}
Here, summation is over reciprocal lattice, $F^{aa}_G(X,X')$ is a form 
factor \cite{Halperin} of two conduction band states at points $X$ and $X'$. 
The absence of the dielectric constant in the Fourier transform of the Coulomb 
potential reflects a lack of lattice screening for these processes. Another 
vertex, which has the same strength $g_q$ but does not contribute to the 
electron-hole pairing, is shown in Fig.~\ref{vertex}c. Several other 
intervalley vertices correspond to processes, when an electron transforms 
into a hole. However, they correspond to higher momentum transfers and we 
neglect them. Generally, the form factors in Eq.~(\ref{gq}) reduce the
matrix element compared to the amplitude of a Coulomb 
potential by a factor of $\sim 2$--5 and can also change the sign of $g_q$. 
Nevertheless, in the absence of necessary numerical band structure results 
we use a somewhat optimistic estimate $|g|\approx 4\pi e^2/Q^2$. 
Returning back to a spatial form of the Schr\"odinger equation we obtain 
a contact-like interaction of excitons in different valleys with a 
strength $g$.

Since the total momentum of excitons vanishes, eigenstates of the 
Schr\"odinger equation are classified according to the irreducible 
representations of the cubic point group $O_h$. Treating 
the intervalley scattering perturbatively, we find 
$E_{e1} = E_e - 2\lambda$ for a nondegenerate exciton state with $A_1$ 
symmetry $\psi_\nu\sim \psi_B(r)(1,1,1)$ and 
$E_{e2} = E_e +\lambda$ for a degenerate doublet with $E$ symmetry: 
$\psi^{(1)}_\nu \sim \psi_B(r)(1,e^{2\pi i/3},e^{4\pi i/3})$ and
$\psi^{(2)}_\nu \sim \psi_B(r)(1,e^{-2\pi i/3},e^{-4\pi i/3})$, 
where $\psi_B(r)$ is a $1s$ hydrogenic wave function. The coupling 
constant is $\lambda = g |\psi_B(0)|^2$ and, therefore, choice of 
the lowest energy exciton depends on the sign of $g$. If interaction 
between excitons is attractive ($g>0$) the symmetric $A_1$-state has 
lower energy. If it is repulsive, then the $E$-doublet is more stable. 
One can easily estimate $\lambda$ using unperturbed $\psi_B(r)$: 
$\lambda = 0.15E_x$. We have also calculated the effect of intervalley 
vertices Figs.~\ref{vertex}b and \ref{vertex}c on the ground state 
energy of the metallic $e$-$h$ liquid shown by dashed line in 
Fig.~\ref{egs}. This correction is much smaller than a change of the 
exciton energy and does not exceed 3\%. Qualitatively, such 
a difference is explained by different orders of the two corrections: 
for the metallic plasma it is a second-order effect, while the shift of
exciton energy is a first order effect. Another effect of lifting 
threefold degeneracy of excitons in different valleys is suppression of 
multi-exciton molecules.

As one can see from Fig.~\ref{egs} the exciton energy is still above 
the ground state energy of the metallic phase, though the two energies 
move closer to each other. However, it appears that our estimate for 
the intervalley scattering effect may be too conservative. The origin of 
extra enhancement is similar to a mechanism of anomalously large 
hyperfine splitting of donor states in Si proposed by Kohn and Luttinger 
\cite{Kohn2}. The actual exciton energy is shifted below prediction of the 
effective mass theory because of both the intervalley scattering effect 
and a so called central cell correction \cite{central}. Contrary to a 
naive point of view, related changes in the amplitude $\psi(0)$ are 
determined by a long distance Schr\"odinger equation rather than by 
an exact short distance $e$-$h$ Hamiltonian. Since the actual $E_e$ is not 
an eigenvalue of the effective mass equation, there is no solution for 
this energy which satisfies both boundary conditions at $r=0,\infty$. 
Therefore, actual exponentially decaying wave-function develops 
a singularity at short distances. We find that for a moderate 20\% shift 
of $E_e$ the probability to find electron near hole $|\psi(0)|^2$ is 
enhanced by a factor of 4 compared to $|\psi_B(0)|^2$. This effect 
significantly increases $\lambda$, especially 
for the $A_1$-singlet \cite{central}. As 
a result, the excitonic level in the hexaborides must lie well below 
the metallic minimum of $E_{\rm g.s.}(r_s)$.

We have considered stability of different phases of the $e$-$h$ liquid 
in the hexaborides. They include a semiconducting state with no carriers, 
a dilute Bose gas of excitons and a dense 
electron-hole liquid. The latter can be clearly distinguished from 
the other two states in infrared optical conductivity by its large Drude 
peak. If an excitonic instability develops in a dense $e$-$h$ liquid, 
$\sigma(\omega)$ must also have an edge-type singularity at the excitonic 
gap. Semiconducting and free exciton gas states, on the other hand, 
have no significant features in $\sigma(\omega)$ because of the absence 
of direct optical transitions between the two bands. Which of the three 
states really occurs depends on the band gap parameter $E_G$. Note, that 
in the scenario (a), see Fig.~\ref{egs(rs)}, a first order transition 
between a semiconducting state and a dense plasma takes place at a 
positive $E_G$, which always corresponds to a semiconductor in a single 
electron picture. Small values of $|E_G|\sim E_x=0.08$~eV found in 
the band structure calculations \cite{Hasegawa,Massida} indicate that 
the hexaborides can be close to an instability of semiconducting state, 
which, as we argued for CaB$_6$, transforms into a dilute exciton gas. 
The other possible instability in the hexaborides, if $E_G$ is varied on  
experiment, is a gas-liquid transition between a dilute Bose gas of 
excitons and a dense $e$-$h$ plasma.

We suggest that some of the above phase transformations could be induced 
in the hexaborides by applying hydrostatic pressure, which changes the 
band gap $E_G$. Applied uniaxial stress can further lift the degeneracy 
between different valleys and, thus, reduce significantly an energy of 
the metallic phase \cite{Brinkman,Rice}. Our main result, 
stability of a dilute exciton gas in CaB$_6$ either at normal conditions 
or under pressure, provides a new way to look at the 
Bose condensation of excitons. 
Another intrinsic mechanism for gap variations can also come from 
impurity doping, in particular from La- substitution for a divalent 
element. Therefore, doped hexaborides can differ from undoped 
compounds not only because of unequal number of electrons and holes,
but also in terms of an appropriate starting picture: 
a Bose gas of excitons or a dense $e$-$h$ liquid. In the latter 
case screening effects must be quite significant due to a high 
density restriction for $e$-$h$ plasma: $r_s\lesssim 1$.

We thank W. Kohn for useful discussions. 
Financial support for this work was provided
by Swiss National Fund.


\begin{thebibliography}{10}
\bibitem{central}
Central cell corrections to the exciton energy come in part because 
of a lack of lattice screening of $e$-$h$ interaction at distances 
smaller than lattice constant. We estimate this effect as 
$\Delta E_e \sim -0.2E_x$ for the $A_1$-state. Note that similar shift is also 
present for the ground state energy of $e$-$h$ plasma with $r_s\sim 1$. 
Hence, central cell correction alone cannot explain stability of 
a dilute exciton gas. The $E$-states have vanishing probability 
to find an electron at a hole and central cell corrections
are much smaller for them.
\end{thebibliography}
\end{document}